\documentclass{PoS}
\usepackage{amsmath,amssymb}
\usepackage{bm}

\def\mc#1{{\mathcal #1}}

\title{Hadronic Interactions from Lattice QCD }

\ShortTitle{Hadronic Interactions from Lattice QCD }

\author{\speaker{Andr\'{e} Walker-Loud}%
        \thanks{I would like to thank my NPLQCD Collaborators who produced the results I am presenting.}\\
       College of William and Mary \\
       E-mail: \email{walkloud@wm.edu}}

\abstract{I will discuss recent results of the NPLQCD Collaboration regarding the calculation of hadronic interactions with lattice QCD.  A particular emphasis will be spent on pi-pi scattering and other meson interactions.}

\FullConference{8th Conference Quark Confinement and the Hadron Spectrum\\
		 September 1-6, 2008\\
		 Mainz. Germany}

\begin{document}

%

%
\section{Two-Particle Energy Levels in a Finite Euclidean Volume}

In a finite Euclidean volume, one does not calculate scattering matrix elements directly~\cite{Maiani:1990ca}, but rather one calculates the volume dependent distortion of the two-particle energy levels due to the particle interactions which, below inelastic thresholds, can be related to the infinite volume scattering phase shift; this is known as L\"{u}scher's method~\cite{Hamber:1983vu}.  For example, for two degenerate particles in a cubic spatial volume of size $L^3$ and periodic boundary conditions, the distortion of the free-particle energy eigenvalues is related to the scattering phase shift through through a regulated three-dimensional summation,
\begin{equation}\label{eq:Sfunc}
\Delta E_{2m} = 2\sqrt{p^2 + m^2} - 2m^2 \quad \longrightarrow \quad
\frac{p \cot \delta(p)}{m} = \frac{1}{m \pi L} S \left(\frac{p^2 L^2}{16\pi^2} \right)
\end{equation}
where $S(z)$ can be found in Refs.~\cite{Beane:2003yx},
\begin{equation}
	S(z) \equiv \sum_\mathbf{n}^{|\mathbf{n}| < \Lambda} \frac{1}{|\mathbf{n}|^2 - z} - 4\pi \Lambda\, .
\end{equation}

%
\section{Precision $I=2\ \pi\pi$ Scattering}

Low energy $\pi\pi$ scattering is theoretically the best understood hadron-hadron interaction.
At leading order in chiral perturbation theory ($\chi$PT)~\cite{Weinberg:1966kf,Gasser:1983yg}, the effective field theory describing the low-energy interactions of QCD, the s-wave scattering lengths are uniquely predicted~\cite{Weinberg:1966kf}.  The subleading corrections are known to one-~\cite{Gasser:1983yg} and two-loops~\cite{Bijnens:1995yn}.  Each of these higher order corrections depends upon unknown low-energy-constants (LECs) which must be determined from comparison with experiment or lattice calculations.

\begin{figure}[b]
\center
\includegraphics[width=0.42\textwidth]{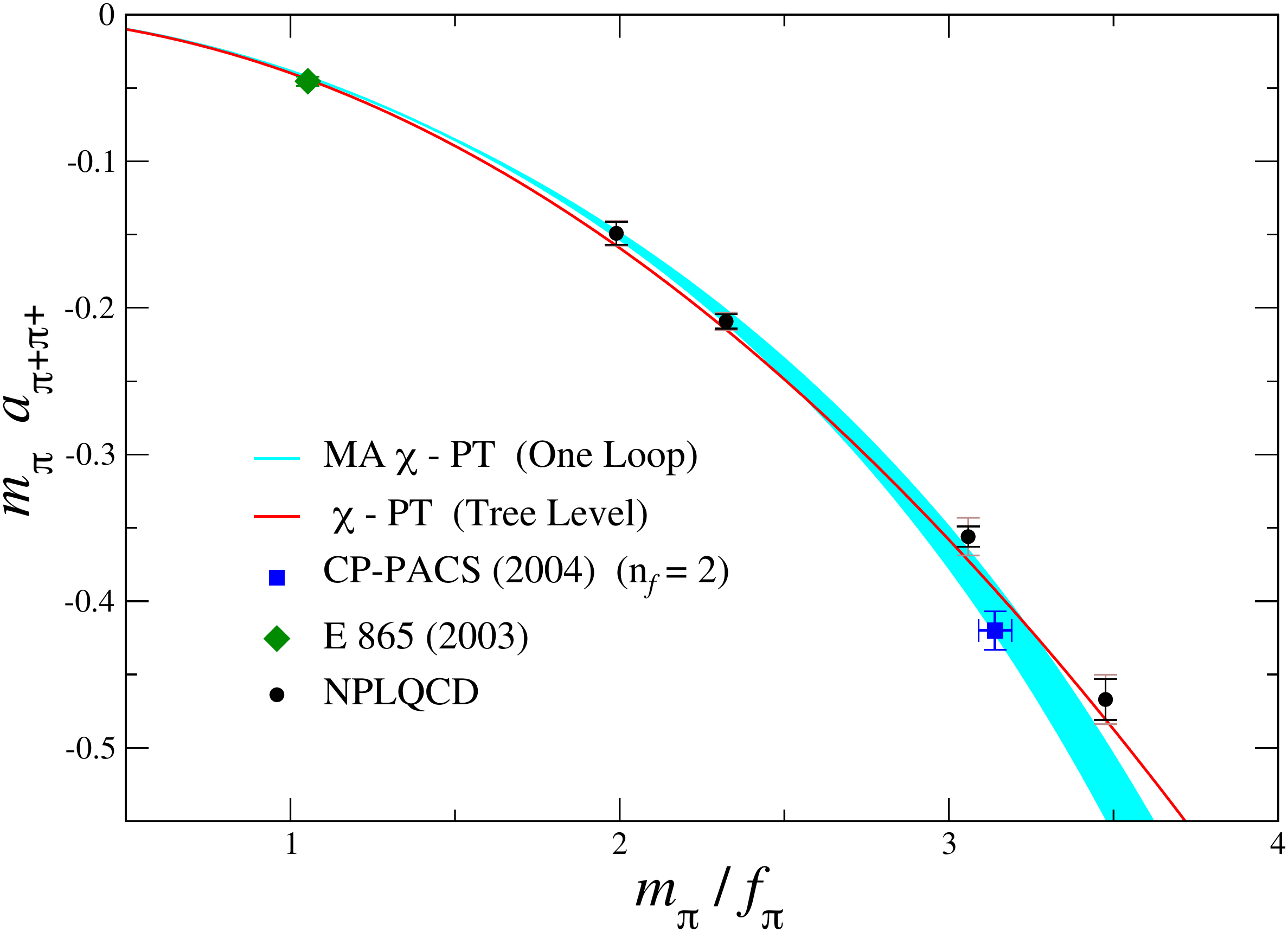}
\caption{\label{fig:mpiapipi} \textit{Chiral extrapolation of $I=2\ \pi\pi$ scattering length using mixed-action $\chi$PT.}}
\end{figure}

Low energy $\pi\pi$ interactions are also numerically very tractable as the pions are the lightest hadrons of QCD.  The exotic $I=2\ \pi\pi$ scattering channel
\footnote{Due to the disconnected quark diagrams, the scalar $I=0$ channel is numerically more expensive and noisier.} %
is therefore an ideal process to study with lattice QCD before exploring the more challenging pion-nucleon and nucleon-nucleon interactions.  The NPLQCD Collaboration performed the first lattice calculation of the $I=2\ \pi\pi$ scattering length with 2+1 dynamical flavors of light quarks ($up$, $down$, and $strange$)~\cite{Beane:2005rj}.  Using continuum $\chi$PT to perform the extrapolation to the charged pion mass, they found $m_\pi a_{\pi\pi}^{I=2} = -0.0426 \pm 0.0019$.  The calculation was performed by computing domain-wall valence quark propagators~\cite{Kaplan:1992bt} on the Asqtad improved~\cite{Orginos:1999cr} coarse MILC lattices~\cite{Bernard:2001av}, the mixed-action lattice scheme of the LHP Collaboration~\cite{WalkerLoud:2008bp}.  Subsequently, a high-precision update of this result has been published~\cite{Beane:2007xs}, finding at the charged pion mass
\begin{equation}
	m_{\pi^+} a_{\pi\pi}^{I=2} = -0.04330 \pm 0.00042\, .
\end{equation}
There are several ingredients which were necessary for this precision calculation; high statistics with the lattice calculation, systematic treatment of the lattice-artefacts from the discretization and finite volume, and chiral symmetry.
%
%
\begingroup
\begin{table}[t]
\caption{\label{tab:MILCcnfs}Parameters of NPLQCD lattice calculation of $I=2\ \pi\pi$ scattering~\cite{Beane:2007xs}, using coarse MILC ensembles~\cite{Bernard:2001av} with lattice spacing $b\sim 0.125~\texttt{fm}$ and volume $L\sim 2.5~\texttt{fm}$.}
\center
\begin{tabular}{c|cccc}
\hline\hline
$\{m_\pi,m_K\}[\texttt{MeV}]$ & \{290,580\} & \{350,595\} & \{490,640\} & \{590,675\} \\
$N_{cfg}\times N_{source}$ & 468\ $\times$\ 16 & 658\ $\times$\ 20 & 486\ $\times$\ 24 & 564\ $\times$\ 8 \\
\hline\hline
\end{tabular}
\end{table}
\endgroup
A series of papers allowed for all of these systematic errors to be addressed in the updated calculation~\cite{Beane:2007xs}.  First, Ref.~\cite{Beane:2007xs} included an order of magnitude increase in statistics compared to the original calculation~\cite{Beane:2005rj}, see Table~\ref{tab:MILCcnfs} for details.  In Refs.~\cite{Chen:2005ab}, the lattice spacing corrections to the scattering length were determined by utilizing the relevant mixed-action $\chi$PT.  In Refs.~\cite{Bedaque:2006yi}, the exponentially suppressed volume corrections to Eq.~\eqref{eq:Sfunc} were determined.%
\footnote{The volume corrections computed in Ref.~\cite{Bedaque:2006yi} were determined with continuum $\chi$PT.  The volume corrections for partially-quenched/mixed-action theories are also known~\cite{PQpipiFV} and were used in Ref.~\cite{Beane:2007xs}.} 
The other important ingredient is chiral symmetry, both the chiral symmetry of the domain-wall valence propagators used in the computation, as well as the fact that the $\pi\pi$ scattering lengths must vanish in the chiral limit.  When expressed in terms of the lattice-physical pion mass and decay constant, the mixed-action extrapolation formula for the scattering length is given at one loop by~\cite{Chen:2005ab}
\begin{equation}
m_\pi a_{\pi\pi}^{I=2} = -(2\pi)\frac{m_\pi^2}{(4\pi f_\pi)^2} \left\{
	1
	+\frac{m_\pi^2}{(4\pi f_\pi)^2} \left[
	3 \ln \left( \frac{m_\pi^2}{\mu^2} \right) -1 -l_{\pi\pi}^{I=2}(\mu)
	-\frac{\tilde{\Delta}_{PQ}^4}{6m_\pi^4} \right] \right\}\, ,
\end{equation}
where $l_{\pi\pi}^{I=2}(\mu)$ is the same linear combination of renormalized Gasser-Leutwyler coefficients as appears in continuum $\chi$PT.  There is no explicit dependence upon the valence-sea mesons~\cite{Orginos:2007tw} and the only explicit sea quark effects are contained in the last term, where for the quark mass tuning employed in Refs.~\cite{Beane:2005rj,Beane:2007xs}, $\tilde{\Delta}_{PQ}^2 = b^2 \Delta_{\mathrm{I}}$, which is the staggered taste-Identity splitting, which has been computed on the coarse MILC lattices~\cite{Aubin:2004fs}.
The resulting chiral extrapolation is presented in Fig.~\ref{fig:mpiapipi}.
One striking feature of this lattice calculation, for the full range of quark masses used, the resulting scattering length is consistent Weinberg's tree level prediction~\cite{Weinberg:1966kf}.  Aside from the quantum corrections to the pion mass and decay constant, there is no evidence of higher order effects.  This feature is not specific to $I=2\ \pi\pi$ scattering but has also been found in computations of both the $I=3/2\ K\pi$ and $I=1\ KK$ scattering lengths~\cite{Beane:2006gj}.

\section{Nuclear and Hyper-Nuclear Interactions}

\begin{figure}[t]
\center
\begin{tabular}{cc}
\includegraphics[width=0.35\textwidth]{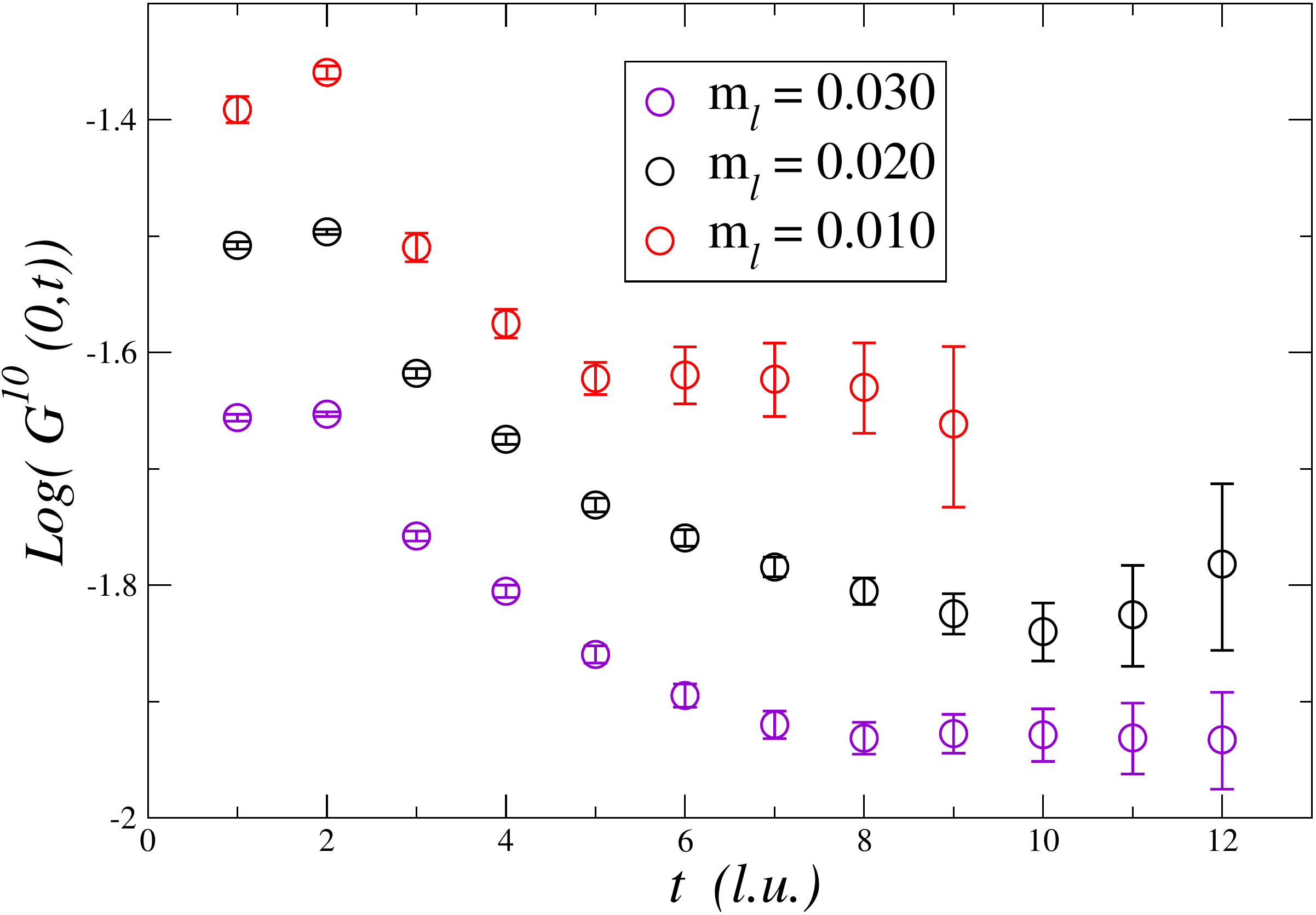}
&
\includegraphics[width=0.35\textwidth]{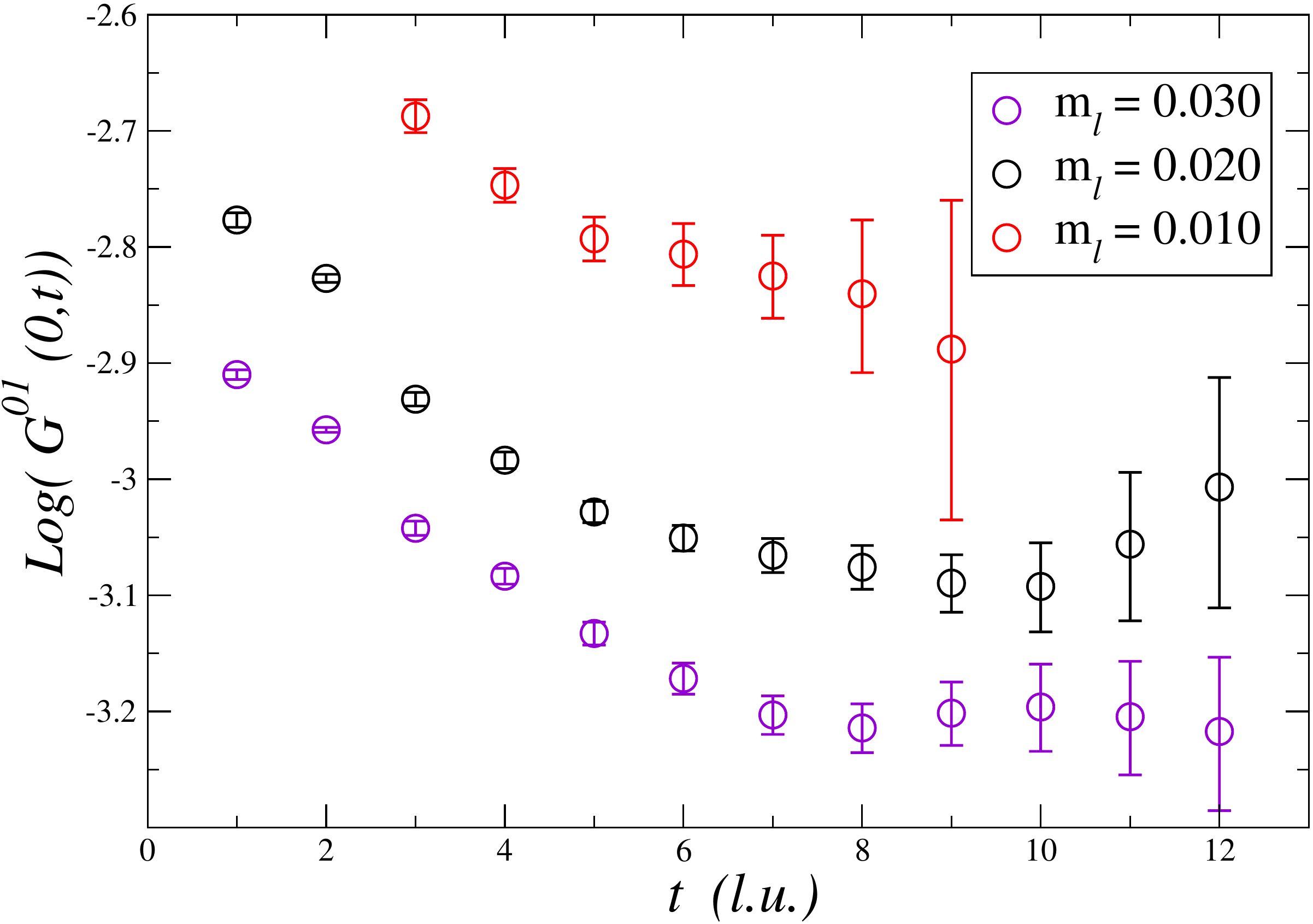}
\end{tabular}
\caption{\label{fig:NN} \textit{${}^1S_0$ and ${}^3S_1$ nucleon-nucleon correlation functions taken from Ref.~\cite{Beane:2006mx}.}}
\end{figure}

The main objective of the NPLQCD Collaboration is to make predictions for the structure and interactions of nuclei using lattice QCD.  To that end, the first dynamical calculations of nucleon-nucleon and hyperon-nucleon scattering lengths were performed recently~\cite{Beane:2006mx}.  However, lattice calculations of (multi)-baryon correlation functions suffer from a exponential degradation of the signal-to-noise~\cite{LePage}, as can be seen in Fig.~\ref{fig:NN}.
This noise degradation is currently the main obstacle to computing multi-baryon correlation functions at light pion masses.  Solutions to this problem will require several orders of magnitude increase in statistics as compared to the calculations in Refs.~\cite{Beane:2006mx}, especially as one utilizes lighter quark masses.  There has been a proposal to alleviate this noise problem with a clever choice of boundary conditions~\cite{Bedaque:2007pe}, although this proposal has yet to be implemented numerically.

%
\section{Kaon Condensation}
In addition to calculating properties of two pions or two nucleons, the NPLQCD Collaboration has utilized these same computing resources to calculate properties of up to twelve pions and kaons~\cite{Beane:2007es,Detmold:2008yn}.  In order to analyze these correlation functions, the energy levels of $n$ non-relativistic identical bosons interacting in a periodic cubic spatial volume of length $L$ was determined through $\mc{O}(L^{-7})$~\cite{Beane:2007qr}.  For example, we were able to compute the kaon chemical potential as a function the density of $K^-$s at densities relevant for neutron star phenomenology.  Additionally, three-pion and three-kaon interactions were determined from these calculations.

%
\section{Conclusions}
The use of lattice field theory methods to compute properties of hadron interactions has rapidly progressed in the last few years~\cite{Beane:2008dv}.  Along the path to computing two-baryon interactions, we have seen a precision calculation of the $I=2\ \pi\pi$ scattering length as well as the computation of systems containing up to 12 pions or kaons, allowing the first prediction for both the three-pion and three-kaon interactions.  In the next few years we will continue to see this rapid progression as the lattice community moves into the era of \texttt{peta-flop} computations.

\end{document}